\documentclass[12pt,thmsa]{article}
\usepackage{sw20lart}

\input{tcilatex}
\begin{document}

\author{Emilio Santos \\
Departamento de F\'{i}sica, Universidad de Cantabria, Santander, Spain}
\title{The arrow of time and the Bell inequalities }
\date{December, 7, 2016 }
\maketitle

\begin{abstract}
It is recalled that closed (isolated) systems are essentially reversible
whilst open systems like the Earth, or living beings on it, are irreversible
because they are not isolated. Earth and life irreversibility derives from
the evolution of the universe, which is a consequence of its special initial
conditions. It is stressed that, although relativity theory forbids that
information travels faster than light, it does not forbid influences of an
event on its past light cone. Therefore the violation of Bell inequalities
in loophole-free experiments is compatible with relativity theory. A
correlation formula, alternative to Bell's, is proposed as the starting
point for hidden variables models fitting in relativity.
\end{abstract}

\section{Introduction}

Several recent experiments have exhibited the loophole-free violation of a
Bell inequality\cite{Hensen},\cite{Shalm},\cite{Giustina}. The result has
been interpreted as the ``death by experiment for local realism'', this
being the hypothesis that ``the world is made up of real stuff, existing in
space and changing only through local interactions ...about the most
intuitive scientific postulate imaginable''\cite{Wiseman}. In this paper I
will argue that the claimed death of local realism requires some refinements.

It is common wisdom that the most celebrated supporter of local realism was
Albert Einstein, whence recalling his views may clarify the subject. His
opinions about realism will not be commented here (see, e. g. \cite{Harrigan}%
), but it is appropriate to comment on his idea about (relativistic)
locality, stated as ``On one supposition we should, in my opinion,
absolutely hold fast: the real factual situation of the system S2 is
independent of what is done with the system S1 , which is spatially
separated from the former.''\cite{Einstein}. This quotation is usually
interpreted as Einstein\'{}s support for ``relativistic causality'', this
used as synonymous of locality. So for instance in the pioneer paper by John
Bell\cite{Bell}. However this interpretation is misleading as explained in
the following.

Causality is commonly viewed as the assumption that the present may
influence the future, but not the past, which in (special) relativity would
mean that an event may be influenced only by events in its past light cone,
that is neither by spacelike separated events nor by events in the future
light cone. However Einstein sentence did not exclude influences by events
in the future light cone. Indeed he was well aware that the laws of physics
do not distinguish future from past, as in the often quoted passage from his
letter of condolences upon the death of his friend Michele Besso: ``Michele
has left this strange world just before me. This is of no importance. For us
convinced physicists the distinction between past, present and future is an
illusion, although a persistent one.''\cite{EinsteinBesso}. Indeed the
concept of temporal causality, stating that an event may influence its
future but not its past, is related to our experience as living beings, but
it is alien to the laws of physics.

The main purpose of this paper is to stress that a (loophole-free) violation
of a Bell inequality does not imply influences between spacelike separated
events provided that we allow influences of the future on the past.

\section{The arrow of time vs. microscopic reversibility}

The name ``arrow of time'' was introduced by Arthur Eddington in 1927. He
wrote ``I shall use the phrase time's arrow to express this one-way property
of time which has no analogue in space''\cite{Eddington}. Thus the arrow of
time refers to the distinction between past and future that we observe in
nature. At present it is used more specifically with reference to the
problem of explaining the irreversibility that we experience, which is not
trivial taking into account that the laws of nature are invariant under time
reversal (except for a small violation in the decay of some elementary
particles like $K$ mesons that will be ignored here). There are many books
and articles devoted (or discussing) the arrow of time and a review is out
of scope of this paper, where I will only discuss a few points that
sometimes have been the source of confusion.

The existence of an arrow of time was formalized by Clausius with the
concept of entropy and its postulated increase for any spontaneous evolution
of an isolated system. The entropy was introduced in physics as a kind of
measure of the ``quality'' of energy. For instance mechanical and
gravitational energy have high quality because they may be transformed
completely in other forms, but this is not the case for heat because only a
part of it can be transformed in work (mechanical energy). In the particular
case of energy transfer taking place exclusively in the form of heat, a
simple quantitative calculation of the entropy change, $\Delta S,$ of a
system is possible, namely 
\begin{equation}
\Delta S=\int \frac{dQ}{T},  \label{2.2}
\end{equation}
$Q$ being the heat entering the system and $T$ the absolute temperature. For
other cases the calculation is more involved. Clausius realized that in the
processes that are possible in the laboratory the total entropy never
decreases. This led to postulate that entropy never decreases in closed
systems, that was the first scientific statement about the existence of an
arrow of time. For instance if we put a hot body in contact with a cold one
the heat goes spontaneously from the former to the latter until they have
equal temperature. This fits in the increase of entropy as is easily derived
from eq.$\left( \ref{2.2}\right) $ leading to 
\[
\Delta S=\int \frac{dQ}{T_{cold}}-\frac{dQ}{T_{hot}}>0, 
\]
which is positive taking into account that $dQ>0$ ($dQ<0)$ is defined as
energy that enters (leaves) the body and obviously $T_{hot}>T_{cold}.$

The fundamental step towards the solution of the apparent contradiction
between the\textit{\ irreversibility of spontaneous (macroscopic) evolution
vs.} \textit{reversibility of the fundamental (microscopic) laws of nature }%
was made by Boltzmann, who gave a microscopic interpretation of entropy.
Boltzmann realized that irreversibility is always associated to macroscopic
systems and he proposed that it is due to the tendency towards more probable
states in the spontaneous evolution. Then Boltzmann introduced a relation
between the entropy, $S$, of a composite system and the number $N$ of
microscopic states of the system that correspond to a given macroscopic
state, that is 
\begin{equation}
S=k_{B}\log N,  \label{2.3}
\end{equation}
where $k_{B}$ is today named Boltzmann constant. A standard example is a box
divided in two equal parts by a wall with a small hole on it, filled with an
amount of gas consisting of $n$ molecules. If we define a microscopic state
by specifying which gas molecules are present in each part of the box, there
is only one state with all molecules in the left (or in the right). In this
state $N=1$ and eq.$\left( \ref{2.3}\right) $ gives $S=0.$ If at time $t=0$
the box starts in this state, after some time $t=T$ there will be several,
say $j,$ molecules on the left and $n-j$ on the right. Hence the number of
microstates equals the number of ways to choose $j$ molecules amongst $n$,
that is 
\[
N=\frac{n!}{j!(n-j)!}>1\Rightarrow S>0. 
\]
The most probable state will correspond to $j=n/2$ whence, 
\[
S_{\max }=k_{B}\log N_{\max }\simeq k_{B}n\log 2. 
\]

Boltzmann's work was one of the great achievements in the history of
physics, but it did not solve the problem of the arrow of time as was soon
pointed out by several authors, in particular Loschmidt and Poincar\'{e}. I
think that in order to clarify the subject it is important to distinguish
between the evolution of systems in experiments made in the laboratory and
whot happens on Earth.

\section{Evolution of closed systems in the laboratory}

I will speak about LAB experiments in a wide sense, including processes
induced by human beings like those of chemical industry. In any case I will
refer only to evolution of isolated systems because it is obvious that
evolution subject to external influences may present irreversibility induced
by them. In the example of the box, commented in the previous section, the
irreversibility is related to 
\[
S(T)>S(0). 
\]
The Loschmidt argument applied to this example is as follows. If the system
was isolated since well before $t=0$ it is the case that at time $t=-T$ the
gas would be filling both parts of the box. In fact the evolution backwards
in time between $t=0$ and $t=-T$ would be identical to the evolution forward
in time between between $t=0$ and $t=T$ with all velocities reversed at time 
$t=0$. Therefore in terms of the entropy we may write 
\[
S(-T)=S(T)>S(0). 
\]
The reversal of velocities is appropriate for classical mechanical systems
consisting particles. In quantum physics the complex conjugation of the
wavefunction is substituted for the velocities reversal.

Any reader will immediatily argue that nobody has ever seen an isolated box
with a quantity of gas having an homogeneous density (say at time $t=-T$) to
evolve spontaneously towards a state with all the gas concentrated in a part
of the box (at time $t=0$). This is true, but the point is that we, human
beings, are able to prepare a box having gas in only one part and then
observe the evolution towards the future, $t=T$, but we are unable to
observe towards the past, $t=-T$, the evolution of an isolated system
prepared at time $t=0.$\textit{\ That is, the irreversibility in the LAB is
not a feature of the material systems themselves, but it derives from our
fundamental irreversibility as living beings. }This irreversibility
constrains us to observe what happens at times $t>0$ to a system prepared by
us at time $t=0$, but we are unable to prepare an isolated system in such a
way that we could observe its evolution towards the past. In section 5 we
shall see that apparently there are experiments where it is possible to
derive the existence of influences ``towards the past'' from actual
experiments.

The conclusion is that closed (isolated) systems are reversible, this being
a straightforward consequence of the reversibility of the fundamental laws
of physics. In particular if a system is isolated between times $-T$ and $T$
and at time $t=0$ it is out of equilibrium, then it will be more close to
equilibrium both at time $T$ and at time $-T$. Of course this does not apply
to the Earth as a whole or to the living beings, including humans, because
they are not isolated. This point will be commented in more detail in the
next section.

\section{The irreversibility of the Earth, the living beings and the
universe.}

Explaining the irreversibility of living beings, including humans, is rather
trivial once we know that the universe is expanding. The universe may be
assumed an isolated system, governed by reversible laws, but its initial
state was very special. In that state it was far from equilibrium and
consequently its evolution has been irreversible. The expansion combined
with the attractive nature of gravity caused that the initial almost
homogeneous plasma evolved giving rise to galaxies and stars. The stars
frequently have associated planets giving rise to solar systems. Every
planet receives energy from its star, this causing irreversible evolution.
Incidentally in a stationary universe the existence of (irreverible) living
beings would be difficult to explain except introducing additional
assumptions.

Our solar systems was formed about 5 billion years ago. After some period
the Earth, initially very hot, became cold arriving at an approximate
stationary state with a separation of the solid crust, the sea and the
atmosphere. In that cold Earth life surged and then evolved until the
appearance of human beings. The evolution in that period has been clearly
irreversible and the reason is obvious. The (stationary) Earth is not an
isolated system. Asides from minor perturbations, the main cause of
irreversibility is the fact that it is receiving energy at high temperatura (%
$T_{in}\simeq 5800K$) from the Sun and sending away a similar power by
radiation at lower temperatura ($T_{out}\simeq 300K$). This produces a net
increase of entropy of the universe at a rate 
\[
\frac{dS}{dt}=\frac{W}{T_{out}}-\frac{W}{T_{in}}>0, 
\]
where $W$ is the average power received from the Sun or emitted by the Earth
to outer space. The irreversibility of Earth is responsible for the
irreversibility of the living beings, including us. That is life in Earth is
an irreversible process because living beings are interacting with the
environment and the process increases the entropy.

In summary all closed (isolated) systems are reversible. However any
macroscopic system that at a given time, say $t=0$, is out of equilibrium
would evolve towards equilibrium both for the past and the future as far as
the system remains isolated. This implies that, if we study the system only
towards de future it will evolve irreversibly approaching equilibrium. This
is the case for the universe as a whole that we can study only \textit{after}
the big bang.

\section{Acausality in Bell experiments}

The consequence of the facts commented in the previous sections is that
locality interpreted as \textit{relativistic (temporal) causality does not
follow from relativity theory} because the theory is time reversal
invariant. Therefore if two events $A$ and $B$ are timelike separated it is
equally correct to say that $A$ is the cause of $B$ or that $B$ is the cause
of $A$. That is the fact that $B$ happens later or earlier than $A$ is
irrelevant. Thus in physics we should speak about correlation between
timelike events rather than causality. In sharp contrast, in biology or
social sciences the concept of causality attached to time ordering is very
relevant, the systems studied by these sciences being essentially open and,
consequently, irreversible.

In a Bell experiment\cite{Shalm},\cite{Giustina} there are two parties,
Alice and Bob, measuring some observable property of one particle each. I
will label $A(B)$ the observable measured by Alice (Bob). Typically $A$ may
be one of two possible photon polarizations and similar for $B$. I shall
label the results of the measurements $a$ and $b$ respectively. Pairs of
particles in an appropriate (entangled) state are produced in the source.
Bell's proposal for the expectation of the product of observables, $%
\left\langle AB\right\rangle ,$ in what he named ``local hidden variables
(LHV) model'', was 
\begin{equation}
\left\langle AB\right\rangle =\int \rho (\lambda )a\left( A,\lambda \right)
b\left( B,\lambda \right) d\lambda ,  \label{4.0}
\end{equation}
where $\lambda $ labels the state produced in the source (typically two
entangled photons), $\rho $ is the probability density of states and $a(b)$
is the result obtained by Alice (resp. Bob), typically $a=1$ (detection) or $%
a=0$ (absence of detection) and similar for $b$. (Bell considered
deterministic LHV models\cite{Bell}, but the generalization to probabilitic
models is straightforward\cite{Santos}). Bell pointed out that the result $a$
should not depend on what Bob is measuring, say $B$, and similarly $b$
should not depend on $A$. In loophole-free tests these conditions are
carefully implemented via performing the measurements by Alice and Bob in
spacelike separated regions. This requirement was strongly supported by
Einstein in the paragraph that we reproduce in the introduction of this paper%
\cite{Einstein}. However Bell also demanded that $\rho $ should not depend
on $A$ or $B$ (neither on $a$ or $b$), the reason being the fact that the
measurements are in the future light cone of the state production on the
source, a condition that Bell included under the concept of locality. In
order to see more clearly how Bell's locality condition agrees with
(relativistic) causality, we may substitute $\sigma \left( \lambda ,\mu
\right) $ for $\rho (\lambda )$ in eq.$\left( \ref{4.0}\right) $, where $\mu 
$ represents all relevant events in the backward light cone with influence
in the state preparation (e.g. the properties of the laser and the nonlinear
crystal where the entangled photon pair is produced). Therefore Bell's
correlation formula eq.$\left( \ref{4.0}\right) $ may be written more
explicitly 
\begin{equation}
\left\langle AB\right\rangle =\int d\lambda \int d\mu \sigma (\lambda ,\mu
)a\left( A,\lambda \right) b\left( B,\lambda \right) .  \label{4.2}
\end{equation}
It is easy to see that eq.$\left( \ref{4.2}\right) $ implies eq.$\left( \ref
{4.0}\right) $ provided that we identify 
\begin{equation}
\int d\mu \sigma (\lambda ,\mu )=\rho (\lambda ).  \label{4.3}
\end{equation}

However \textit{influences from the forward light cone are not forbidden by
relativity theory}. Thus we should substitute 
\[
\left\langle AB\right\rangle =\int d\lambda \int d\mu \sigma (\lambda ,\mu
,a,b)a\left( A,\lambda \right) b(B,\lambda ) 
\]
for eq.$\left( \ref{4.2}\right) ,$ thus including the possible influence of
the most relevant events in the future of the state preparation, namely the
absoption, or not, of the corresponding photon by Alice or Bob. With the
identification eq.$\left( \ref{4.3}\right) $ this becomes 
\begin{equation}
\left\langle AB\right\rangle =\int \rho (\lambda ,a,b)a\left( A,\lambda
\right) b\left( B,\lambda \right) d\lambda ,  \label{4.1}
\end{equation}
rather than eq.$\left( \ref{4.0}\right) ,$ as appropriate for models of
correlation. It may be interpreted saying that the probability of the state
in the source depends on whether the photons will be detected or not, which
of course depends on what measurement are to perform Alice and Bob, this
being governed by the results of two independent random generators\cite
{Shalm},\cite{Giustina}. In actual experiments the state created in the
source is spacelike separated from both random generations and these are
spatially separated from each other. However both the state production in
the source and Alice\'{}s random generation are in the past light cone of
Alice measurement, and similar for Bob. Hence eq.$\left( \ref{4.1}\right) $
is consistent with no influences between spacelike separated events, which
should be the real meaning of locality.

The experiments\cite{Shalm},\cite{Giustina} have refuted eq.$\left( \ref{4.0}%
\right) $ because they have violated its consequence, namely the Bell
inequality. In sharp contrast a Bell inequality cannot be derived from eq.$%
\left( \ref{4.1}\right) $. Therefore the theoretical arguments provided in
this paper show that the empirical evidence support the thesis that eq.$%
\left( \ref{4.1}\right) $ rather than eq.$\left( \ref{4.0}\right) $ is the
correct starting point to understand correlations, including quantum
correlations associated to entanglement. Consequently eq.$\left( \ref{4.1}%
\right) $ should be the basis for hidden variables models consistent with
relativity theory.

Many people are aware of the fact that the (loophole-free) violation of a
Bell inequality seems to create a conflict with relativity theory. The most
popular scapes to this conclusion are the following \cite{Brunner}. Some
authors simply reject the need (or even the possibility) of hidden variables
models. For other people the solution is more sophisticated, they
distinguish superluminal influences from superluminal signals and assume
that only the latter are forbidden by relativity theory. Indeed superluminal
signals are also forbidden by quantum mechanics (no-signalling theorem).
Other solutions less popular are the absolute determinism or the assumption
that some (causal) common influence correlates the random generations with
the system preparation in the source. The latter would amount to assume that 
$\lambda $ is correlated with $A$ and/or $B$ due to some events in the
common backward light cone, a possibility certainly compatible with
relativity but more implausible than eq.$\left( \ref{4.1}\right) $ in my
opinion.

In conclusion I propose that the loophole-free violation of the Bell
inequality should be interpreted as showing that an event may influence
other events on its \textit{past} light cone, whence eq.$\left( \ref{4.1}%
\right) $, rather than the more restrictive eq.$\left( \ref{4.0}\right) ,$
should be the basis for hidden variables models compatible with relativity$.$
Eq.$\left( \ref{4.1}\right) $ might be interpreted in ``human language''
saying that the system in the source ``knows'' in advance whether every
photon will be ``later'' detected or not. This statement sounds rather
counterintuitive, but it fits in relativity theory. In contrast suggesting
that influences may travel with superluminal speed may sound less
counterintuitive, but in my opinion violates relativity theory.

An interpretation of quantum mechanics that takes into account the possible
influence of the future on the past has been proposed with the name \textit{%
transactional interpretation }\cite{Cramer}. The relation of that
interpretation with the proposal made here will not be discussed further in
this paper.

\end{document}